
\font\mybb=msbm10 at 12pt
\def\bb#1{\hbox{\mybb#1}}

\def\bE {\bb{E}}

\def\bM {\bb{M}}


\tolerance=10000
\input phyzzx
\input epsf.tex
 \def\unit{\hbox to 3.3pt{\hskip1.3pt \vrule height 7pt width 
.4pt \hskip.7pt
\vrule height 7.85pt width .4pt \kern-2.4pt
\hrulefill \kern-3pt
\raise 4pt\hbox{\char'40}}}


\REF\pkta{P.K. Townsend, Phys. Lett. {\bf 350B} (1995) 184.}
\REF\wita{E. Witten, Nucl. Phys. {\bf B443} (1995) 85.}
\REF\hs{G.T. Horowitz and A. Strominger, Nucl. Phys. {\bf B360} (1991) 197.}
\REF\pol{J. Polchinski, Phys. Rev. Lett. {\bf 75} (1995) 4724.}
\REF\sen{A. Sen, Phys. Rev. {\bf D54} (1996) 2964.}
\REF\doug{M. Douglas, D. Kabat, P. Pouliot and S.H. Shenker, {\it D-branes and
Short Distances in String Theory}, hep-th/9608024.}
\REF\douglas{M. Douglas, {\it Branes within Branes}, hep-th/9512077.}
\REF\bgprt{E. Bergshoeff, M. de Roo, M.B. Green, G. Papadopoulos and P.K.
Townsend, Nucl. Phys. {\bf B470} (1996) 113.}
\REF\polch{J. Polchinski, S. Chaudhuri and C. Johnson, {\it Notes on D-branes},
hep-th/9602052.}
\REF\gg{M.B. Green and M. Gutperle, Nucl. Phys. {\bf B476} (1996) 484.}
\REF\BBJ{K. Behrndt, E. Bergshoeff and B. Janssen, {\it Intersecting D-branes
in Ten and Six Dimensions}, hep-th/9604168.}
\REF\memdy{J.M. Izquierdo, N.D. Lambert and G. Papadopoulos and P.K. Townsend,
Nucl. Phys. {\bf B460}, (1996) 560.}
\REF\gp{G. Papadopoulos, {\it A brief guide to p-branes}, hep-th/9604068.}
\REF\glpt{M.B. Green, N.D. Lambert, G. Papadopoulos and P.K. Townsend, Phys.
Lett. {\bf 384B} (1996) 86.}
\REF\tsey{A. Tseylin and J. Russo, {\it Waves, boosted branes and BPS states
in M Theory}, hep-th/9611047.}
\REF\guv{R. G{\" u}ven, Phys. Lett. {\bf 276B} (1992) 49.}
\REF\paptown{G. Papadopoulos and P.K. Townsend, Phys. Lett. {\bf 380B} (1996)
273.}
\REF\ark{A. Tseytlin, Nucl. Phys. {\bf B475} (1996) 149.}
\REF\kleb{A. Tseytlin and I. Klebanov, Nucl. Phys. {\bf B475} (1996) 179.}
\REF\jer{J.P. Gauntlett, D.A. Kastor and J. Traschen, Nucl. Phys. 
{\bf B478} (1996)
544.}
\REF\bho{E. Bergshoeff, C.M. Hull and T. Ort\'\i n, Nucl. Phys.
{\bf B451} (1995) 547.}


\Pubnum{ \vbox{ \hbox{R/96/43} } }
\pubtype{}
\date{November 1996}

\titlepage

\title {\bf Kaluza-Klein on the Brane\foot{Revised version}}

\author{G. Papadopoulos and P.K. Townsend}
\address{DAMTP, University of Cambridge,
\break
Silver St., Cambridge CB3 9EW, U.K.}

\abstract{The M-theory interpretation of certain D=10 IIA p-branes 
implies the existence of worldvolume Kaluza-Klein modes which are expected to
appear as 0-brane/p-brane bound states preserving 1/4 of the spacetime
supersymmetry. We construct the corresponding solutions of the effective
supergravity theory for $p=1,4$, and show that no such solution exists for
$p=8$.}

\endpage


\chapter{Introduction}

There is now ample evidence that the IIA superstring theory is an $S^1$
compactification of an 11-dimensional supersymmetric quantum theory called
M-theory. It was pointed out in [\pkta,\wita] that this interpretation requires
the presence in the non-perturbative IIA superstring theory of 
BPS-saturated particle states carrying Ramond-Ramond (RR) charge,
corresponding to the Kaluza-Klein (KK) modes of D=11 supergravity, and it was
argued that these should be identified with the IIA 0-branes. At the time, the
only evidence for the required 0-branes was the existence of extreme electric
`black hole' solutions of the effective IIA supergravity theory [\hs], 
but their
presence in the IIA superstring theory was subsequently confirmed by the
interpretation of D-branes as the carriers of RR charge [\pol]. 

Actually, one needs not just the D-0-brane, for which the effective field 
theory
realization is the extreme black hole of lowest charge associated with the 
first
KK  harmonic, but also a bound state at threshold in the system of $n$ 
D-0-branes
for each $n>1$, a prediction that has still to be confirmed although there is
good evidence that it is true [\sen]. Assuming that these bound states exist,
M-theory provides a KK interpretation of the D-0-branes of IIA superstring
theory. However, as emphasized in [\pkta], {\it all} the IIA p-branes must have
a D=11 interpretation. Indeed, many of them can be interpreted as reductions,
either `direct' or `double', of D=11 branes, i.e. M-branes. The cases of
interest to us here are those IIA p-branes that that have a D=11 interpretation
as (p+1)-branes wrapped around the compact 11th dimension. The massless
worldvolume action for the D=10 p-brane is then a dimensional reduction 
on $S^1$
of the worldvolume action of the (p+1)-brane of M-theory. Thus, the D=11
interpretation of these D=10 p-branes requires the existence of massive
particle-like excitations `on the brane' that can be identified with the KK
harmonics of the `hidden' $S^1$. From the D=10 string theory perspective these
excitations can only be BPS-saturated 0-brane/p-brane bound states [\doug].
Moreover, since the worldvolume KK states preserve 1/2 of the (p+1)-dimensional
worldvolume supersymmetry and the p-brane preserves 1/2 of the spacetime
supersymmetry, these `brane within brane' states must preserve 1/4 of the
spacetime supersymmetry [\douglas]. The IIA $p$-branes for which we 
should expect to find such bound states are (i) the 1-brane, i.e. the 
fundamental IIA
superstring, since this is a wrapped D=11 membrane, (ii) the D-4-brane, since
this is a wrapped D=11 fivebrane, and possibly (iii) the D-8-brane, since
it has been suggested [\bgprt] that the D-8-brane might be a wrapped 
D=11 ninebrane.

The required bound states are not difficult to identify in case (i). 
A fundamental string can end on a 0-brane; actually, charge conservation 
requires a 0-brane to be the end of at least two fundamental strings. Two 
such strings can be joined at their other ends to produce a closed string 
loop with a 0-brane `bead'. One could replace the 0-brane by a bound state 
of several 0-branes. Thus, the bound states needed for the KK interpretation of the IIA superstring as a wrapped D=11 membrane are an immediate consequence 
of the 0-brane bound states needed for the KK interpretation of the 
effective IIA supergravity theory. This is not so in cases (ii) and 
(iii) for which we need 
to find bound states of D-0-branes with D-4-branes or D-8-branes. The 
existence of such bound states is consistent with the `D-brane 
intersection rules' [\polch,\gg] which allow, in particular, the possibility of a p-brane within a q-brane preserving 1/4 of the supersymmetry for $p=q$ mod 4.
The issue of 0-brane/4-brane bound states has been discussed recently in the
context of the D-brane effective action [\doug]. Here we investigate this
question in the context of solutions of the effective IIA supergravity theory. 
We shall show that solutions representing 0-branes within p-branes preserving
precisely 1/4 of the supersymmetry exist for $p=1,4$ but not otherwise
(completing previous partial constructions by other methods [\BBJ]). 

This result is consistent with the standard D=11 interpretation of all the type
II p-branes for $p\le6$ but {\it not} with the interpretation of the
IIA 8-brane as an $S^1$-wrapped M-theory ninebrane. A further argument
against the M-theory ninebrane interpretation of the IIA 8-brane comes from
consideration of a solution of IIA supergravity preserving 1/4 of the spacetime
supersymmetry that represents a D-4-brane within a D-8-brane (the metric for
this solution is already known [\BBJ]; here we present the complete solution). 
If there were an M-theory ninebrane it would be natural to interpret this
4-brane within 8-brane solution as a fivebrane within a ninebrane wrapped on
$S^1$ along a fivebrane direction. However, if such an M-theory configuration
were to exist it could also be reduced to a IIA 5-brane within an 8-brane
but there does not exist any such solution of IIA supergravity preserving 
precisely 1/4 of the spacetime supersymmetry. 

In what follows we shall use the notation $(q|q,p)$ to represent a q-brane 
within a p-brane preserving 1/4 of the supersymmetry. This is the special 
case of $(r|p,q)$, which we use to denote a solution representing an r-brane
intersection of a p-brane with a q-brane. Thus, in this notation the
supersymmetric solutions representing a 0-brane within a p-brane for $p=1$
and $p=4$ are $(0|0,1)$ and $(0|0,4)$. These solutions have magnetic duals, 
$(5|5,6)$ and $(2|2,6)$, respectively, whose existence is required by
M-theory. To see this recall that the D=11 interpretation of the 6-brane is
simply as a D=11 spacetime of the form $H_4\times \bM_7$ where $H_4$ is a
particular (non-compact)  hyper-K{\" a}hler manifold and $\bM_7$ is
7-dimensional Minkowski spacetime [\pkta]. Clearly, there is nothing to prevent
the worldvolumes of either the D=11 membrane or fivebrane from lying within the
$\bM_7$ factor, and from the D=10 perspective this is a membrane or a 5-brane
within a 6-brane. We shall show how the $(5|5,6)$ and $(2|2,6)$ can also be
deduced from known intersecting M-brane solutions.


\chapter{Branes within branes in IIA supergravity}

As just explained, M-theory predicts the existence of a variety of IIA
supergravity solutions preserving precisely 1/4 of the N=2 spacetime 
supersymmetry that represent `branes within branes' (by `precisely` we 
mean to exclude solutions preserving more than 1/4 of the supersymmetry). 
A summary of these predictions is as follows: we expect $(0|0,p)$ solutions for p=1,4 and possibly p=8, {\it but not otherwise}. We also expect the magnetic 
duals of $(0|0,p)$ for $p=1,4$, and a $(4|4,8)$ solution. 

That there are no $(0|0,2)$, $(0|0,5)$ or $(0|0,6)$ solutions of IIA
supergravity preserving 1/4 (as against any other fraction) of the 
supersymmetry
follows from consideration of the projection operators associated with
Killing spinors. A single p-brane solution is associated with a projection
operator $P_p$, of which precisely half the eigenvalues vanish, such that only
spinors $\kappa$ satisfying $P_p\kappa=\kappa$ can be Killing. This accounts 
for
the fact that such solutions preserve half the supersymmetry. Configurations
representing a p-brane within a q-brane for $p\ne q$ can also preserve some
supersymmetry since $P_p$ and $P_q$ must {\it either} commute {\it or}
anticommute. If $P_p$ and $P_q$ commute then the product $P_pP_q$ is also a 
projector. In such cases one may find a supersymmetric solution preserving 1/4
of the supersymmetry, representing either two intersecting branes or a `brane
within a brane'. If $P_p$ and $P_q$ anticommute then the matrix
$$
\alpha P_p +\beta P_q \qquad (\alpha^2 +\beta^2 =1)
\eqn\newa
$$
is another projector with precisely half of its eigenvalues vanishing. In this
case one can hope to find `brane within brane' solutions preserving 1/2 the
supersymmetry. An example of such a solution is the D=11 membrane within a
fivebrane solution [\memdy]; as shown in [\gp,\glpt], this reduces to a 
$(2|2,4)$ solution of IIA supergravity preserving 1/2 the supersymmetry.
Consideration of T-duality then implies the existence of $(0|0,2)$ solutions
preserving 1/2 the supersymmetry \foot{It has been pointed out to us
independently by J. Maldacena and J. Polchinski that such a solution could be
interpreted as a D-2-brane boosted in the 11th dimension. The solution has
since been constructed [\tsey].}. Here we are interested in solutions
preserving precisely 1/4 of the supersymmetry, so only those cases for which
$P_p$ and
$P_q$ {\it commute} are relevant. When both branes are D-branes one can show
that $P_p$ and
$P_q$ commute if and only if $q=p\ $ mod 4, so that $(0|0,2)$ and $(0|0,6)$
solutions preserving 1/4 of the supersymmetry are immediately excluded, whereas
$(0|0,4)$, $(2|2,6)$ and
$(4|4,8)$ are allowed, as is $(0|0,8)$. This D-brane rule says nothing about
$(0|0,1)$ or $(0|0,5)$ since neither the IIA string nor the IIA
5-brane is a D-brane. It happens that $P_1$ commutes with $P_0$ whereas $P_5$
does not, so a $(0|0,1)$ solution preserving 1/4 of the supersymmetry is 
allowed
whereas a $(0|0,5)$ solution is not. A putative $(5|5,8)$ solution 
preserving 1/4 of the supersymmetry is similarly ruled out. For the 
reason given earlier, this fact is evidence against the existence of 
a $(0|0,8)$ solution. Thus,
the projection operator analysis provides arguments both for and 
against the possibility of a $(0|0,8)$ solution.

Leaving aside $(0|0,8)$, we have now seen that the solutions not expected from 
M-theory considerations are indeed absent, while the solutions that 
M-theory requires to exist are permitted. We shall now show that all of 
the latter, among those mentioned above, not only exist but can be 
constructed from known intersecting M-brane solutions preserving 
1/4 of the supersymmetry [\guv,\paptown,\ark,\kleb,\jer] by means of 
the various dualities connecting M-theory with the IIA and IIB 
superstring theories. The relevant M-theory solutions can be obtained from the
`M-theory intersection rules' determining the allowed M-brane intersections
together with the `harmonic function rule' that allows one to write down the
general solution. For example, the $(0|0,1)$ solution of IIA 
supergravity can be
deduced from the solution of D=11 supergravity associated with the 
intersection
of two membranes  at a point, i.e. $(0|2,2)_M$. This is achieved by
consideration of the `duality chain'
$$
(0|2,2)_M\rightarrow (0|1,2){\buildrel  T \over\rightarrow}
(0|1,1_D)_B{\buildrel  T \over\rightarrow} (0|0,1)\ ,
\eqn\branea
$$
where the subscript $B$ indicates a solution of IIB supergravity and $1_D$
denotes the IIB D-string. In the first step one of the two D=11
membranes is wrapped around the 11th dimension; the corresponding D=10 solution
being obtained by double-dimensional reduction. In the second step we T-dualize
along a direction parallel to the IIA 2-brane to arrive at the IIB solution.
A further T-dualization along one of the two directions determined by the
D-strings leads to the required IIA solution. 

To make clear the unambiguous nature of the derivation we shall give all the
intermediate solutions for this example, while giving just the final result
for the examples to follow. Thus, we begin with the
$(0|2,2)$ solution of D=11 supergravity
$$
\eqalign{ds^2&=U^{1/3} V^{1/3}\big[ -U^{-1} V^{-1} dt^2+ U^{-1}
ds^2(\bE^2)+V^{-1} ds^2(\bE^2)+ds^2(\bE^6)\big]
\cr
G_4&=-3 dt \wedge d\big( U^{-1}J_1+V^{-1} J_2\big)\ ,}
\eqn\bone
$$
where (in the terminology of [\paptown]) $U,V$ are harmonic functions of the
overall transverse space $\bE^6$ and $J_1\oplus J_2$ is a complex structure on
the relative transverse space $\bE^2\oplus \bE^2$ . Double-dimensional 
reduction
along one of the relative transverse directions results in the following 
$(0|1,2)$ solution of IIA supergravity:
$$
\eqalign{
ds_{(10)}^2&= V^{1/2}\big[ -U^{-1} V^{-1} dt^2+ U^{-1}
dx^2+V^{-1} ds^2(\bE^2)+ds^2(\bE^6)\big]
\cr
e^{{4\over3}\phi}&= U^{-{2\over3}} V^{1\over3}
\cr
F_4&=-3 dt \wedge d\big(V^{-1} J_2\big)
\cr
F_3&=-3 dt\wedge dx\wedge d U^{-1}\ , }
\eqn\btwo
$$
where $x$ is the string coordinate.  Next, using the T-duality rules of [\bho]
(adapted to our conventions) to T-dualize  along one of the directions of the
2-brane, we get the $(0|1,1_D)_B$ solution
$$
\eqalign{
ds_{(10)}^2&= V^{1/2}\big[ -U^{-1} V^{-1} dt^2+ U^{-1}
dx^2+V^{-1} du^2 + ds^2(\bE^7)\big]
\cr
e^{{2\over3}\varphi}&=U^{-{1\over3}} V^{1\over3}
\cr
F^{(2)}_3&= -3 dt\wedge du\wedge dV^{-1}
\cr
F^{(1)}_3&= -3 dt\wedge dx\wedge d U^{-1}\ , }
\eqn\bthree
$$
where $\varphi$ is the IIB dilaton.  Finally, we transform \bthree\ using
T-duality along the $u$ direction to get the following $(0|1,0)\equiv (0|0,1)$
solution of IIA supergravity: 
$$
\eqalign{
ds_{(10)}^2&= V^{1/2}\big[ -U^{-1} V^{-1} dt^2+ U^{-1}
dx^2+ds^2(\bE^8)\big]
\cr
e^{{2\over3}\phi}&=U^{-{1\over3}} V^{1\over2}
\cr
F_3&=-3 dt\wedge dx\wedge d U^{-1}
\cr
F_2&= -{9\over2} dt\wedge dV^{-1}\ . }
\eqn\bfour
$$

The above solutions, as others given below, depend on two
{\it independent} harmonic functions, each of which is associated with a single
p-brane. For simplicity, let us suppose that each harmonic function has 
just one singularity (at the position of the brane). Clearly, one must 
further suppose that
both harmonic functions have their singularities at the {\it same} location in
order to be able to interpret the configuration as a `brane within brane'
solution associated with the long range fields of a bound state. The same
solution could equally well represent the simple coincidence of two branes; the
fact that solutions exist with two independent harmonic functions indicates
that any bound state would be a bound state at threshold. It is a
weakness of the effective field theory approach that it cannot distinguish
between a bound state at threshold of two branes or their simple coincidence
because both have the same long range fields. The evidence for bound states
provided by the effective field theory is, therefore, not particularly strong,
Nevertheless, when the harmonic functions in \bfour\ are restricted in the
way just described these solutions do give the long range fields of the 
KK modes
that arise from the wrapping of the D=11 membrane on $S^1$ to give a D=10 
string.

Before proceeding we pause to remark that the magnetic dual of the $(0|0,1)$
solution can be found from the $(3|5,5)_M$ solution of M-theory by the 
following
duality chain:
$$
(3|5,5)_M\rightarrow (3|5,4){\buildrel  T \over\rightarrow} 
(4|5,5_D)_B{\buildrel  T \over\rightarrow} (5|5,6)\ ,
\eqn\bseven
$$
where $5_D$ denotes the D-5-brane of the IIB theory ($5$ denoting the NS-NS
5-brane). In the second step we have T-dualized in a direction parallel to the
IIA 5-brane, which is mapped to the IIB NS-NS 5-brane under this operation. 
The final $(5|5,6)$ solution dual to $(0|0,1)$ (which has been found 
previously
by other means [\ark]), is
$$
\eqalign{ds^2&= U V^{{1\over2}}\big[U^{-1} V^{-1} ds^2(\bM^6)+V^{-1}
dv^2+ds^2(\bE^3)\big]
\cr
e^{{2\over3}\phi}&=U^{1\over3} V^{-{1\over2}}
\cr
F_3&=9 dv \wedge \star dU
\cr
F_2&=27 \star dV\ ,}
\eqn\beight
$$
where $U,V$ are harmonic functions on the Euclidean transverse space 
$\bE^3$ and
$\star$ is the Hodge dual for $\bE^3$.

We turn next to the $(0|0,4)$ case. This can be found from the following
duality chain,
$$
(0|2,2)_M\rightarrow (0|2,2){\buildrel  T \over\rightarrow} 
(0|1_D,3)_B{\buildrel  T \over\rightarrow} (0|0,4)\ , 
\eqn\bafive
$$
where the first step is the direct reduction to D=10 of the D=11 solution. 
The resulting $(0|0,4)$ solution is
$$
\eqalign{ds^2&=U^{{1\over2}} V^{{1\over2}} \big[ -U^{-1} V^{-1} dt^2+V^{-1}
ds^2(\bE^4)+ds^2(\bE^5)\big]
\cr
e^{{2\over3}\phi}&=V^{1\over2} U^{-{1\over6}}
\cr
F_4&=3\star dU
\cr
F_2&=-{9\over2} dt\wedge dV^{-1}\ ,}
\eqn\bfive
$$
where $U,V$ are harmonic functions on $\bE^5$ and $\star$ is now the Hodge dual
for $\bE^5$. The magnetic dual of this solution is $(2|2,6)$, which can 
be found
from the duality chain
$$
(3|5,5)_M\rightarrow (2|4,4){\buildrel  T \over\rightarrow} 
(2|3,5_D)_B{\buildrel  T \over\rightarrow} (2|2,6)\ .
\eqn\bten
$$
The final $(2|2,6)$ solution is
$$
\eqalign{ds^2&=U^{{1\over2}}V^{{1\over2}}\big[U^{-1} V^{-1}
ds^2(\bM^3)+U^{-1}ds^2(\bE^4)+ds^2(\bE^3)\big]
\cr
e^{{2\over3}\phi}&=V^{1\over6} U^{-{1\over2}}
\cr
F_4&=-{3\over2}\epsilon(\bM^3)dV^{-1}
\cr
F_2&=27\star dU \ ,}
\eqn\bnine
$$
where $U,V$ are harmonic functions on $\bE^3$ and $\star$ is the Hodge dual for
$\bE^3$.  

We remark that both $(0|0,4)$ and its magnetic dual $(2|2,6)$ can also be found 
from $(1|2,5)_M$ as follows:
$$
(1|2,5)_M\rightarrow (1|2,4){\buildrel  T \over\rightarrow} 
(0|1_D,3)_B{\buildrel  T \over\rightarrow} (0|0,4)\ ,
\eqn\bsix
$$
and
$$
(1|2,5)_M\rightarrow (1|2,4){\buildrel  T \over\rightarrow} 
(1|1_D,5_D)_B{\buildrel  T \over\rightarrow} (2|2,6)\ .
\eqn\beleven
$$

The above solutions confirm the current D=11 interpretations of all IIA
p-branes for $p\le6$. In addition, the duality chain of \beleven\ can be
continued as follows:
$$
(2|2,6) {\buildrel  T \over\rightarrow} (3|3,7)_B  {\buildrel  T
\over\rightarrow} (4|4,8) \ .
$$
In principle, the 7-brane appearing in the penultimate solution is the
D-7-brane. However, the 7-brane solution needed for this construction is the
`circularly-symmetric' 7-brane of IIB supergravity since, as shown in [\bgprt],
it is this solution that is mapped to either the 6-brane or the 8-brane
solution of $S^1$ compactified IIA supergravity. A further point is that the
T-duality transformations to be used in the last link of the duality chain are
the `massive' ones of [\bgprt] connecting solutions of IIB supergravity with
those of the {\it massive} IIA supergravity theory. Apart from these subtleties
the construction proceeds as before, with the final result
$$
\eqalign{ds^2&=U^{1\over2}V^{1\over2}\big(U^{-1}V^{-1}ds^2(\bM^5)+
U^{-1}ds^2(\bE^4)+ dy^2\big)
\cr
e^{{2\over3}\phi}&=V^{-{1\over 6}} U^{-{5\over6}}
\cr
M&=\partial_yU
\cr
F_4&=3 \epsilon(\bE^4)\partial_yV\ ,}
\eqn\ctwo
$$
where $U,V$ are harmonic functions of $y$. 

Finally, we return to the question of whether there exists a $(0|0,8)$ solution
which might represent KK modes in a possible M-theory ninebrane 
interpretation of the IIA 8-brane. If it exists we should be able 
to deduce it from
M-theory. It cannot be so deduced from the intersecting M-brane solutions
considered so far, but there exists a solution of D=11 supergravity preserving
1/4 of the supersymmetry that has been interpreted as the intersection of two
M-theory fivebranes on a string, i.e. as a $(1|5,5)$ solution [\jer]. Taking 
this solution as the starting point of the following duality chain
$$
\eqalign{
(1|5,5)_M &\rightarrow (0|4,4) {\buildrel T \over\rightarrow}  (0|3,5_D)_B
{\buildrel  T \over\rightarrow} (0|2,6) \cr
 & {\buildrel  T \over\rightarrow} (0|1_D,7)_B {\buildrel T \over\rightarrow}
(0|0,8)\ , }
\eqn\branetwo
$$
we could apparently deduce the existence of the sought $(0|0,8)$ solution.
However, the starting $(1|5,5)$ solution has a quite different form from the
other intersecting M-brane solutions. In particular, the two harmonic functions
associated with each fivebrane are independent of the `overall transverse'
coordinate. On the other hand, consistency with the KK ansatz needed for the
various T-duality steps in the above chain requires that both harmonic
functions be independent of all the other coordinates. Therefore, the only
acceptable starting solution for the above duality chain is the special case
of the $(1|5,5)$ solution for which both harmonic functions are constant;
this is just the Minkowski vacuum which obviously preserves all the
supersymmetry rather than just 1/4 of it. Of course, this shows only that a 
IIA $(0|0,8)$ solution preserving precisely 1/4 of the supersymmetry 
cannot be obtained from a particular
starting point. However, supposing such a solution to exist we could reverse
the steps in the duality chain \branetwo\ to deduce the existence of a
$(1|5,5)$ solution of `conventional' form for intersecting M-branes, i.e. with
both harmonic functions depending only on the overall transverse coordinate.
It is not difficult to see that there is no such solution because the
associated 4-form field strength does not satisfy the field equation
$d*G=G\wedge G$. Thus, there is no $(0|0,8)$ solution of the required type.


\chapter{Conclusions}

The interpretation of certain p-brane solutions of IIA supergravity
as wrapped (p+1)-branes of M-theory requires the
existence of massive KK modes `on the brane'. In turn, this requires the
existence (and in other cases, absence) of `brane within brane' solutions of
IIA supergravity preserving 1/4 of the supersymmetry. We have shown that the
list of such solutions is compatible both with the current M-theory
interpretations of the IIA p-branes with $p\le6$, but not with an
interpretation of the IIA 8-brane as an M-theory 9-brane. 

\vskip 0.5cm

\noindent{\bf Acknowledgments:}  GP thanks The Royal Society for a  University
Research Fellowship. We thank the organisers of the Benasque Centre for Physics
in Spain, where part of this work was done, Michael Douglas,
Michael Green and Christopher Hull for discussions. We also thank Juan
Maldacena and Joseph Polchinski for comments on an earlier version of this
paper and especially Eric Bergshoeff, who pointed out a serious error in it. 

\refout
\end